\documentclass[aps,prl,twocolumn,amsmath,superscriptaddress]{revtex4-1}  
\usepackage{bbm}
\usepackage{mathrsfs}
\usepackage{amsmath}
\usepackage{amsfonts}
\usepackage[colorlinks=true,citecolor=blue,anchorcolor=blue]{hyperref}
\usepackage{graphicx,epstopdf}
\usepackage{float}
\usepackage{subfigure}
\usepackage{epsfig}
\usepackage{dcolumn}
\usepackage{bm}
\usepackage{color}
\usepackage{natbib}
\usepackage{amssymb}
\usepackage{xcolor}
\usepackage{braket}
\begin{document}
\title{Direct Observation of Unidirectional Density Wave and Band splitting in a Single-Domain Trilayer Nickelate Pr$_4$Ni$_3$O$_{10}$}

\author{Zhicheng Jiang}
%%\email{jiang_zc@ustc.edu.cn}
\thanks{Equal contributions}
\affiliation{National Synchrotron Radiation Laboratory and School of Nuclear Science and Technology, University of Science and Technology of China, Hefei, 230026, China}

\author{Enkang Zhang}
%%\email{23210190049@m.fudan.edu.cn}
\thanks{Equal contributions}
\affiliation{State Key Laboratory of Surface Physics and Department of Physics, Fudan University, Shanghai, 200433, China}
\affiliation{Shanghai Research Center for Quantum Sciences, Shanghai, 201315, China}

\author{Yuxin Wang}
%%\email{wangyuxin@ucas.ac.cn}
\thanks{Equal contributions}
\affiliation{Kavli Institute for Theoretical Sciences, University of Chinese Academy of Sciences, Beijing, 100190, China}

\author{Zhengtai Liu}
\email{liuzt@sari.ac.cn}
\affiliation{Shanghai Synchrotron Radiation Facility, Shanghai Advanced Research Institute, Chinese Academy of Sciences, Shanghai 201210, China}
\affiliation{Shanghai Institute of Microsystem and Information Technology, Chinese Academy of Sciences, Shanghai 200050, China}

\author{Jishan Liu}
\affiliation{Shanghai Synchrotron Radiation Facility, Shanghai Advanced Research Institute, Chinese Academy of Sciences, Shanghai 201210, China}
\affiliation{Shanghai Institute of Microsystem and Information Technology, Chinese Academy of Sciences, Shanghai 200050, China}

\author{Runfeng Zhang}
\affiliation{National Synchrotron Radiation Laboratory and School of Nuclear Science and Technology, University of Science and Technology of China, Hefei, 230026, China}

\author{Xinnuo Zhang}
\affiliation{National Synchrotron Radiation Laboratory and School of Nuclear Science and Technology, University of Science and Technology of China, Hefei, 230026, China}

\author{Wenchuan Jing}
\affiliation{Shanghai Institute of Microsystem and Information Technology, Chinese Academy of Sciences, Shanghai 200050, China}

\author{Yu Huang}
\affiliation{Shanghai Institute of Microsystem and Information Technology, Chinese Academy of Sciences, Shanghai 200050, China}

\author{Qi Jiang}
\affiliation{Shanghai Synchrotron Radiation Facility, Shanghai Advanced Research Institute, Chinese Academy of Sciences, Shanghai 201210, China}
\affiliation{Shanghai Institute of Microsystem and Information Technology, Chinese Academy of Sciences, Shanghai 200050, China}

\author{Mao Ye}
\affiliation{Shanghai Synchrotron Radiation Facility, Shanghai Advanced Research Institute, Chinese Academy of Sciences, Shanghai 201210, China}
\affiliation{Shanghai Institute of Microsystem and Information Technology, Chinese Academy of Sciences, Shanghai 200050, China}

\author{Kun Jiang}
\email{jiangkun@iphy.ac.cn}
\affiliation{Institute of Physics, Chinese Academy of Sciences, Beijing National Laboratory for Condensed Matter Physics, Beijing, 100190, China}

\author{Jun Zhao}
\email{zhaoj@fudan.edu.cn}
\affiliation{State Key Laboratory of Surface Physics and Department of Physics, Fudan University, Shanghai, 200433, China}
\affiliation{Shanghai Research Center for Quantum Sciences, Shanghai, 201315, China}
\affiliation{Institute of Nanoelectronics and Quantum Cumputiing, Fudan University, Shanghai, 200433, China}
\affiliation{Hefei National Laboratory, Hefei, 230088, China}

\author{Dawei Shen}
\email{dwshen@ustc.edu.cn}
\affiliation{National Synchrotron Radiation Laboratory and School of Nuclear Science and Technology, University of Science and Technology of China, Hefei, 230026, China}

\author{Donglai Feng}
\email{dlfeng@ustc.edu.cn}
\affiliation{New Cornerstone Science Laboratory, University of Science and Technology of China, Hefei, 230026, China}

\begin{abstract}
Unraveling the interplay between density-wave (DW) instabilities and multi-orbital physics is critical for understanding superconductivity in Ruddlesden-Popper nickelates, yet intrinsic electronic features have been persistently obscured by material inhomogeneity and thus the multi-domain averaging effect. Here, we employ micro-focused angle-resolved photoemission spectroscopy ($\mu$-ARPES) on single-domain Pr$_4$Ni$_3$O$_{10}$ to disentangle the complex hierarchy of intrinsic and back-folded bands, explicitly identifying the electronic states driving the DW phase transition. We provide decisive spectroscopic evidence that the low-energy reconstruction is governed by inter-orbital nesting between the $\alpha$ and $\beta$ bands. Specifically, we resolve a orbital-dependent gap of $\sim44$ meV on the $\alpha$ pocket, a value quantitatively consistent with prior measurements, unifying previously conflicting experimental reports regarding the locus and magnitude of the DW gap. Furthermore, we reveal strong orbital-selective mass renormalization in the $d_{z^2}$ states and successfully resolve the long-sought intrinsic trilayer $\beta$-band splitting, establishing a critical lower bound for the outer-layer hopping. These results define a coherent microscopic fingerprint for the trilayer nickelates, identifying the specific nesting channels and correlation effects that underpin the phase diagram.
\end{abstract}

\maketitle
%----------------------------introduction--------------------------------
\section{introduction}
A hallmark of unconventional superconductivity, as exemplified by cuprates and iron-based superconductors, is its close proximity and competition with spin/charge density-wave (DW) orders~\cite{armitage2010progress,dai2015antiferromagnetic}. These DW instabilities are believed to play a critical role in shaping the superconducting ground state, potentially even driving the pairing mechanism itself~\cite{chang2012direct}. Recently, the discovery of high temperature superconductivity in pressurized Ruddlesden–Popper (RP) nickelates, most notably the bilayer La$_3$Ni$_2$O$_7$ and trilayer La$_4$Ni$_3$O$_{10}$, has ignited intense interest in exploring analogs in nickel-based systems~\cite{sun2023signatures,zhu2024superconductivity,zhang2024high,wang2024bulk,ko2025signatures,zhou2025ambient,zhang2025bulk,shi2025pressure}. Similar to their iron-based counterpart, these layered nickelates exhibit robust DW transitions at ambient pressure, which are suppressed under applied pressure, coinciding with the emergence of superconductivity~\cite{wang2024pressure,zhao2025pressure,xu2025collapse}. Deciphering the microscopic electronic structure underpinning the DW phases is therefore crucial to understanding the mechanism of superconductivity in nickelates. 

However, experimentally accessing this electronic structure has proven exceptionally difficult. The intrinsic material complexities, including oxygen non-stoichiometry, structural inhomogeneity~\cite{dong2024visualization,dong2025interstitial,puphal2024unconventional,chen2024polymorphism}, and the coexistence of multiple DW domains~\cite{2024arXiv241006602Z, gupta2025anisotropic}, severely hinder precise spectroscopic characterization. Consequently, while scanning tunneling microscopy (STM) and optical spectroscopy have detected local charge modulations and hints of DW gap formation~\cite{li2025direct,xu2025collapse}, angle-resolved photoemission spectroscopy (ARPES) has yet to unambiguously identify DW-related electronic features. In La$_3$Ni$_2$O$_7$, for instance, clear DW gaps on the Fermi surface remain elusive~\cite{yang2024orbital}. The situation is even more perplexing in La$_4$Ni$_3$O$_{10}$, where conflicting observations have emerged: one study identified an SDW gap on the flat, hole-like $\gamma$ pocket~\cite{li2017fermiology}, whereas a separate study observed the gap on the electron-like $\alpha$ pocket, reporting a size much smaller than that discovered in STM and optical spectroscopy~\cite{yang2024orbital,xu2025collapse}. This experimental ambiguity stands in stark contrast to theoretical simulations, which identify Fermi-surface nesting as a key driver for DW instability, yet direct verification of the expected band folding and gap opening remains elusive~\cite{jia2025lattice}. Compounding this issue, previous ARPES measurements revealed surprisingly similar Fermi surfaces for both trilayer and bilayer nickelates, contradicting the distinct topologies predicted by first-principles calculations~\cite{jiang2024high,chen2024trilayer,yang2024effective,tian2024effective}.

The recently discovered Pr$_4$Ni$_3$O$_{10}$ offers a superior platform due to its high crystalline quality and structural uniformity~\cite{zhang2025bulk}. In this work, we employ micro-focused ARPES ($\mu$-ARPES) to selectively probe a single structural/electronic domain of Pr$_4$Ni$_3$O$_{10}$, effectively circumventing the spectral blurring arising from domain superposition. This domain-resolved approach allows us to directly observe the intrinsic trilayer-splitting of bands, alongside distinct back-folded bands originating from structural distortions and intertwined density-wave orders. By disentangling these complex features in momentum space, we recognize the gap opening related to the SDW. Our results successfully clarify the long-standing ambiguities in the field, providing compelling evidence that the gap opening is driven by a unidirectional, incommensurate SDW order.

%-------------------Main Text-------------------------
\section{Crystal Structure}
\begin{figure}[htbp]
\centering
\includegraphics[width=7.5cm]{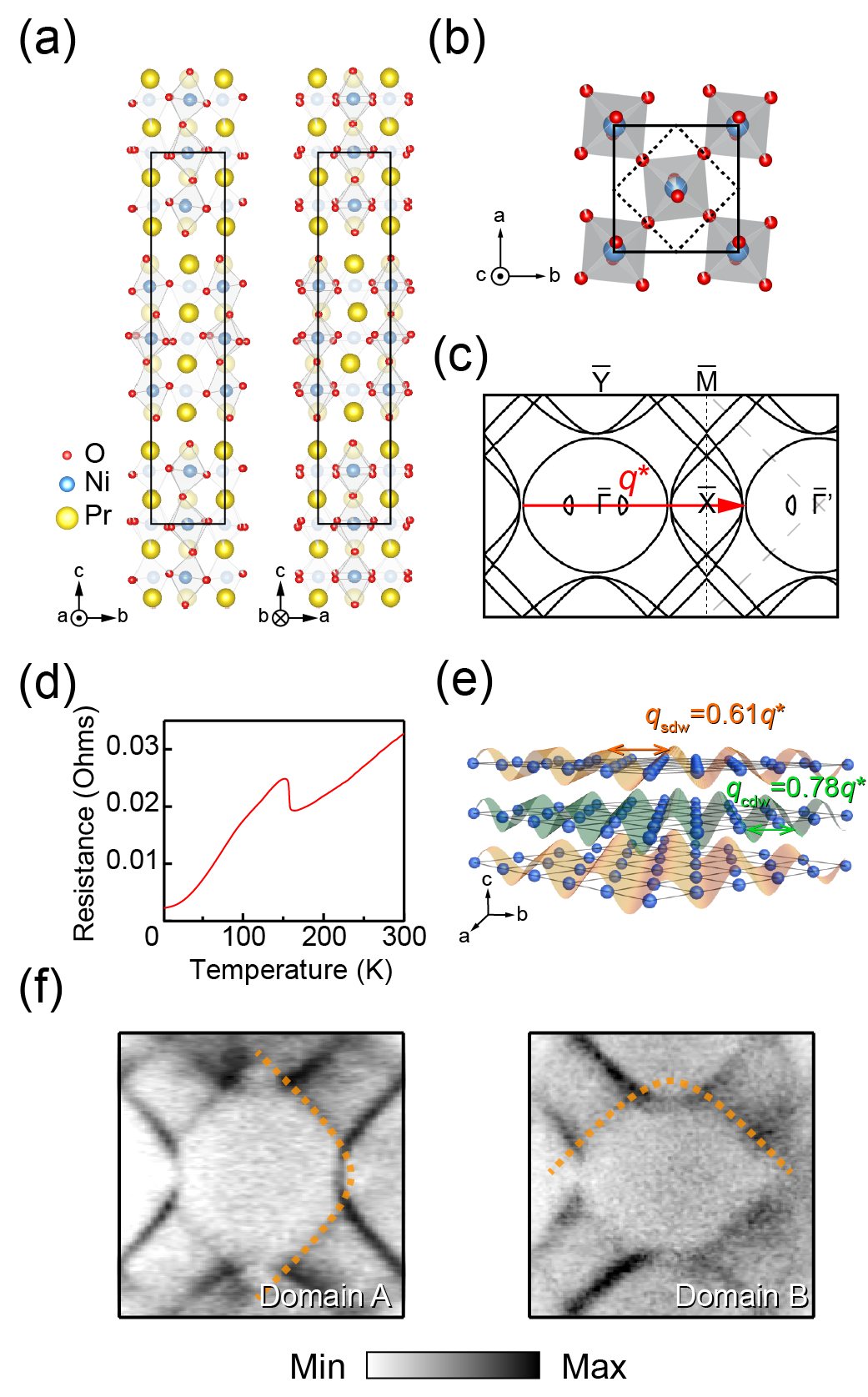}
\caption{
(a) Side view of the crystal structure of Pr$_4$Ni$_3$O$_{10}$, dark frame indicates the unit cell (consisting of two Pr$_4$Ni$_3$O$_{10}$ formula units);
(b) Top view of the crystal structure with a two-dimensional lattice of O and Ni atoms. The dash dark line frame represent the original unit cell without considering the tilted Ni-O octahedra and the solid line frame represents the real structural unit cell by considering the tilting of the Ni-O octahedra;
(c) Calculated Fermi surface of Pr$_4$Ni$_3$O$_{10}$ obtained from the DFT calculations, appended red arrow represent the scattering $\textbf{q*}$ is the unit scattering vector. To note, the anisotropy lattice parameter generate 0.01 \AA$^{-1}$ difference on $k_x$/$k_y$ direction of the Brillouin zone. Therefore the magnitude of unit scattering vector $\textbf{q*}$ can be denoted as $q*~=~\frac{2\pi}{(a+b)/2}$.
(d) In-plane resistivity ($\rho_{ab}$) of Pr$_4$Ni$_3$O$_{10}$;
(e) Schematic SDW and CDW alien with Ni lattice model, the SDW/CDW propagates unidirectionally along the a/b-axis of the crystal lattice.
(f) The two domains with orthogonal electronic structures measured in the Pr$_4$Ni$_3$O$_{10}$.
}
\label{fig1}
\end{figure}

Single crystals of Pr$_4$Ni$_3$O$_{10}$ adopt the $n$ = 3 member of the RP structure series, characterized by perovskite blocks of three NiO$_6$ octahedra layers separated by rock-salt layers [Fig.~\ref{fig1}(a)]. Analogous to its isostructural counterpart La$_4$Ni$_3$O$_{10}$, collective tilting and rotation of these NiO$_6$ octahedra in Pr$_4$Ni$_3$O$_{10}$ at ambient pressure drive a distortion from the tetragonal structure, resulting in an in-plane $\sqrt{2}$ $\times$ $\sqrt{2}$ superlattice reconstruction. Fig.~\ref{fig1}(b) presents a top-down schematic of this structure with Pr atoms omitted for clarity, where the original unit cell is marked by a dashed box and the superlattice cell by a solid frame. Correspondingly, X-ray diffraction measurements confirm that Pr$_4$Ni$_3$O$_{10}$ crystallizes in the monoclinic space group P2$_1$/c, with lattice parameters  $a$ = 5.377 \AA~and $b$ = 5.465 \AA, respectively~\cite{zhang2025bulk}. 

The trilayer structure results in two crystallographically inequivalent NiO$_6$ planes, corresponding to the inner and outer layers. The DFT calculations predict that this structural inequivalence lifts the degeneracy of the $\beta$ bands, which are primarily of Ni $3d_{x^2-y^2}$ character, leading to a sizable band splitting as schematically illustrated in Fig.~\ref{fig1}(c). The resistivity curve measured at ambient pressure [Fig.~\ref{fig1}(d)] exhibits a distinct anomaly at $T_{DW}$ $\approx$ 157 K, marking a transition into a state with coexisting spin- and charge-density-wave (SDW/CDW) order. 

Previous X-ray and neutron scattering results~\cite{jia2025lattice,samarakoon2023bootstrapped} have established that these density waves propagate along a unidirectional diagonal direction of the Ni-square lattice ($\textbf{q*} \parallel a/b$), with $\textbf{q}_{sdw} = 0.61\textbf{q*}$ and $\textbf{q}_{cdw} = 0.78\textbf{q*}$, as shown in Fig.~\ref{fig1}(e). While recent local probes, such as STM on the sibling compound La$_4$Ni$_3$O$_{10}$~\cite{li2025direct}, reported an anisotropic CDW wave vector, this anisotropy has proven elusive to spatially-averaged techniques like conventional photoemission and X-ray scattering~\cite{li2017fermiology,zhou2024revealing}. This failure is widely attributed to spatial averaging over multiple coexisting crystalline or electronic domains.

Our $\mu$-ARPES results, acquired from a single-domain region using a $\sim$ 15 $\times$ 15 $\mu$m$^2$ beam spot, are displayed in Fig.~\ref{fig1}(f). The measured Fermi surface maps deviate significantly from previous ARPES reports~\cite{li2017fermiology,du2025dichotomy}. Specifically, near the Brillouin zone center, we observed a prominent ``arc-like'' feature (marked by orange dashed lines). Although this feature was previously attributed to the $\alpha$ electron pocket, this assignment is inconsistent with the relatively isotropic $\alpha$-band from DFT calculations~\cite{li2017fermiology}. Instead, our data in Fig.~\ref{fig1}(f) clearly show that this feature intersects and hybridizes with the $\beta$ band. In the following sections, we will disentangle and analyze these band characteristics in detail. 

\begin{figure*}[htbp]
\centering
\includegraphics[width=16cm]{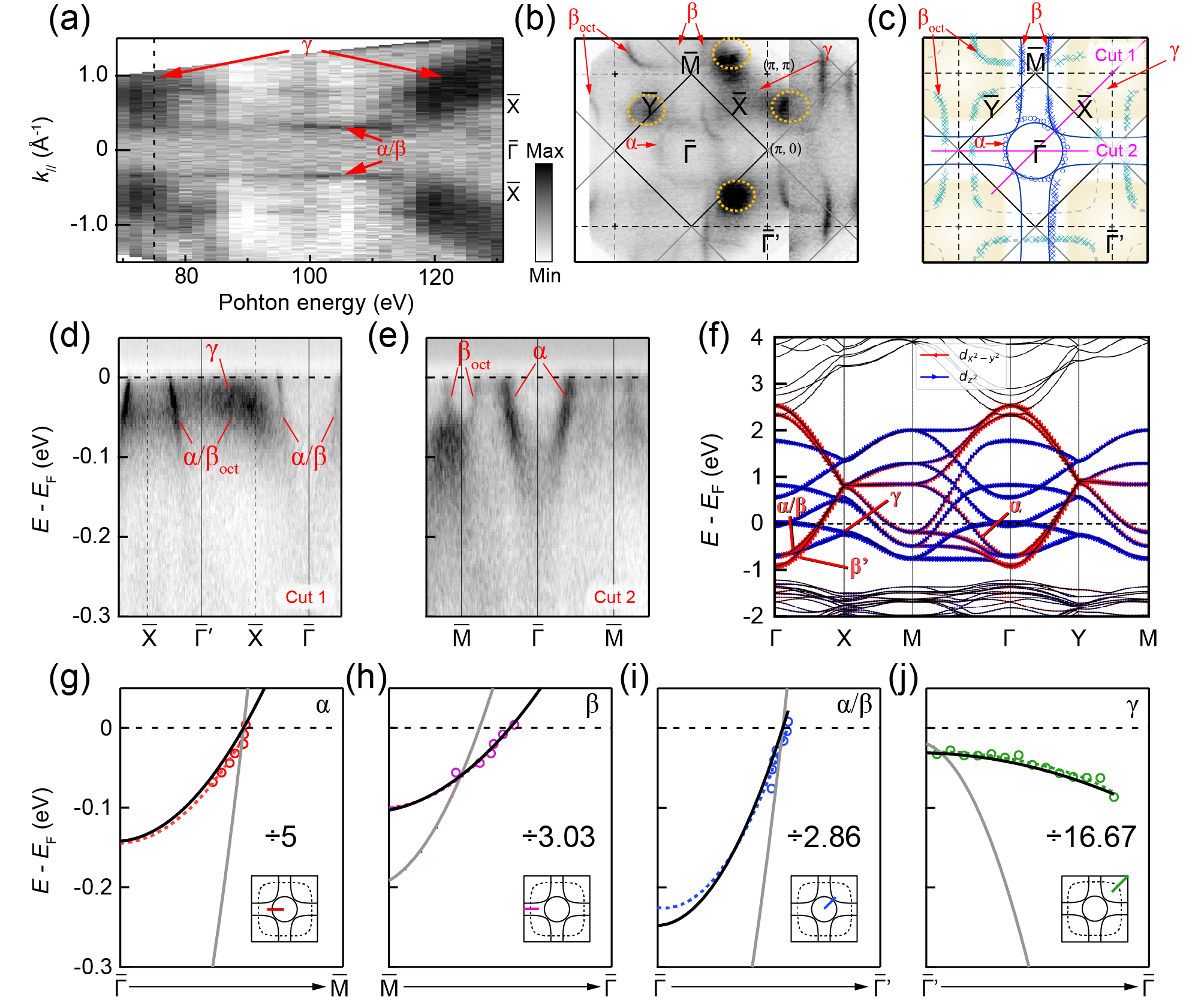}
\caption{
(a) Photon energies-$k_{//}$ mapping along the (0, 0)-($\pi$, $\pi$) direction. It is obtained by integrating the spectral intensity within 50 meV with respect to the Fermi level;
(b) Fermi surface mapping measured at 8~K by using synchrotron-based ARPES with a photon energy of 75~eV. It is obtained by integrating the spectral intensity within 50 meV with respect to the Fermi level. Brillouin zones are appended, with the dark solid lines represent the BZ of pristine lattice, and dark dash lines represent the BZ of octahedral lattice.
(c) Measured Fermi surface of Pr$_4$Ni$_3$O$_{10}$ obtained form (b), appended with momentum cuts marks. It consists of two main Fermi surface sheets, $\alpha$ and $\beta$, and their octahedral back-folded bands $\alpha_{oct}$ and $\beta_{oct}$.
(d-e) Band dispersions measured along momentum cuts Cut1 and Cut2, respectively. The location of the momentum cuts is shown in (c). The observed bands are labeled by their corresponding Fermi surface sheets;
(f) DFT calculated band dispersion of Pr$_4$Ni$_3$O$_{10}$ without considering U. Red lines represent the contribution of Ni 3$d_{x^2-y^2}$ orbitals and blue lines represent the contribution of Ni 3$d_{z^2}$ orbitals.
(g-j) Measured band dipsersions (empty circles) and the corresponding calculated bands (solid lines). The direction of the corresponding momentum cut is shown in the inset at the bottom-right corner of each figure. The gray curve in the figure is the quadratic curve obtained by fitting the original DFT calculation results, while the black curve is the result after being scaled by the corresponding mass enhancement values to match the experimental data.
}
\label{fig2}
\end{figure*}

\section{Intrinsic Electronic Structure and Orbital-Selective Correlations}

Fig.~\ref{fig2}(a) presents a photon-energy-dependent map of the band structure in an intentionally selected single domain along the $\overline{\Gamma}-\overline{X}$ high-symmetry direction. All primary electronic bands can be observed along this cut, with the $\alpha/\beta$ bands located near the Brillouin zone center ($k_{//}$=0) and the higher-intensity $\gamma$ band situated around the $\overline{X}$ points. The $\alpha$/$\beta$ bands exhibit negligible dispersion as a function of photon energy, indicating their two-dimensional character. In contrast, the $\gamma$ band shows pronounced warping, which signifies a substantial dispersion along the out-of-plane momentum direction ($k_z$). These findings are consistent with the primary orbital characters of the bands: the $\alpha$/$\beta$ bands derive mainly from the in-plane Ni $3d_{x^2-y^2}$ orbitals, conferring their 2D nature, while the $\gamma$ band originates from the out-of-plane Ni $3d_{z^2}$ orbital, giving rise to its substantial $k_z$ dispersion.

A photon energy of 75 eV was then chosen for detailed measurements, as it provides the sharpest spectral features, as well as facilitating a direct comparison with previous ARPES studies on RP nickelates~\cite{yang2024orbital,li2017fermiology,du2025dichotomy,au2025universal}. Fig.~\ref{fig2}(b) displays the Fermi surface measured at this photon energy, which clearly resolves the $\alpha$/$\beta$ Fermi pockets (marked by red arrows). Additionally, replica bands from octahedral distortion, denoted as $\beta_{oct}$, can be also clearly resolved. This feature likely arises from the band folding induced by the superlattice potential of the tilted NiO$_6$ octahedra. We then quantitatively extracted the Fermi momenta for all observed pockets through both energy distribution curves (EDC) and momentum distribution curves (MDC) analysis, which are summarized as the contours in Fig.~\ref{fig2}(c). This Fermi surface is highly consistent with that of La$_4$Ni$_3$O$_{10}$~\cite{li2017fermiology,du2025dichotomy}.

Fig.~\ref{fig2}(d) and ~\ref{fig2}(e) present the detailed band dispersions along the high-symmetry directions indicated in Fig.~\ref{fig2}(c). By comparing these data with our orbital-projected first-principles calculations [Fig.~\ref{fig2}(f)], we could unambiguously assign the orbital characters of the observed bands. The $\alpha$ and $\beta$ bands possess predominantly $d_{x^2-y^2}$ character, while the $\gamma$ band is primarily of $d_{z^2}$ character. The $\gamma$ band top is located at $\approx$ 30 meV below $E_F$, similar to that reported for La$_3$Ni$_2$O$_7$~\cite{yang2024orbital}. However, a significant discrepancy emerges: the experimental bandwidths are substantially compressed relative to the DFT calculations. This indicates strong carrier mass renormalization (\textit{i.e.}, self-energy effects) driven by electron correlations in Pr$_4$Ni$_3$O$_{10}$. 

To quantify this, we extracted the mass renormalization factor ($m^*/m_b$) for each band. The $d_{x^2-y^2}$-derived bands show moderate renormalization: $m^*/m_b$ $\approx$ 5 for the $\alpha$ band and $m^*/m_b$ $\approx$ 2.9–3.0 for the $\beta$ band. Strikingly, the $d_{z^2}$-derived $\gamma$ band is renormalized by a factor of $m^*/m_b$ $\approx$ 16.7. This enormous value, while vastly exceeding that of the $d_{x^2-y^2}$ bands, is consistent with the physics of related nickelates. For instance, in bilayer La$_3$Ni$_2$O$_{7}$, the $d_{z^2}$-derived band ($m^*/m_b$ $\approx$ 5.5 $\sim$ 8.3) is also substantially more renormalized than its $d_{x^2-y^2}$ counterparts~\cite{yang2024orbital}. We note that a precise quantification for the $\gamma$ band is complicated by hybridization effects and the opening of the density wave gap near $E_F$. Nevertheless, the enormous relative difference in renormalization is unambiguous. These results provide clear experimental evidence that electron correlations in Pr$_4$Ni$_3$O$_{10}$ are highly orbital-selective, being dramatically stronger for the $d_{z^2}$ states than for the $d_{x^2-y^2}$ states.

\section{Disentangling the DW-Folded Electronic Structure}

\begin{figure*}[htbp]
\centering
\includegraphics[width=18cm]{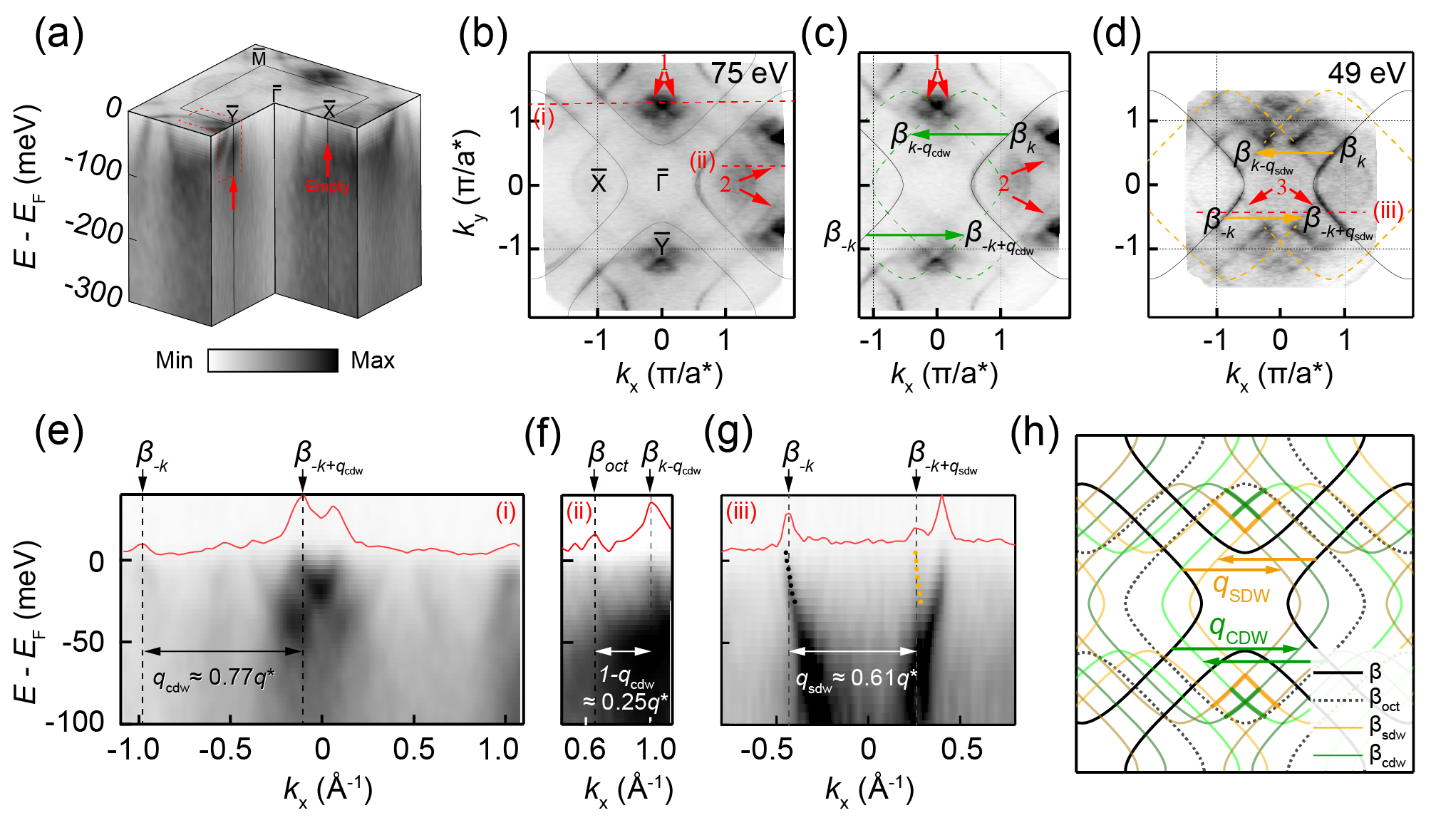}
\caption{
(a) A 3D intensity plot of the ARPES spectra along orthogonal directions $\overline{\Gamma}$-$\overline{X}/\overline{Y}$, with red arrows label the anisotropic features.
(b) Fermi surface mapping measured at 8~K by using 75~eV photons, appended with $\beta$ and $\beta_{oct}$ pockets. It is obtained by integrating the spectral intensity within 10 meV with respect to the Fermi level. The a* = (a+b)/2. The red arrows with label ``1'' mark the crossed extra features circled by $\beta$ and $\beta_{oct}$ pockets, and those labeled by ``2'' mark those extra features parallel to $\beta_{oct}$ pockets.
(c) Fermi surface mapping taken with 75~eV appended with $\beta$ (dark solid frame) and $\beta_{CDW}$ (green solid frame).
(d) Fermi surface mapping measured at 8~K by using 49~eV photons, appended with $\beta$ (dark solid frame) and $\beta_{SDW}$ pockets (orange solid frame);
(e) Band dispersions extracted from the 75~eV mapping along momentum cuts along $k_y$ = 0.775 \AA$^{-1}$ [cut (i), the inset of (e)]; 
(f) Band dispersions extracted from the 75~eV mapping along momentum cuts along $k_y$ = 0.205 \AA$^{-1}$ [cut (ii), the inset of (e)]; 
In (e-f) The MDC (red curve) extracted at Fermi level is appended on the top empty region above the $E_F$. The double-head arrows indicate the scattering vector;
(g) Band dispersions extracted from the 49~eV mapping along momentum cuts along $k_y$ = -0.21 \AA$^{-1}$ (labeled by red dash line in the inset);
(h) Schematic Fermi surface combined with all the pristine $\beta$ (dark solid) and back-folded $\beta$ result from octahedral tilting superlattice (dark dash lines), SDW (origin solid lines) and CDW (green solid lines).
}
\label{fig3}
\end{figure*}

In addition to the primary $\alpha$ and $\beta$ bands and their superlattice-induced replicas, our $\mu$-ARPES measurements reveal faint spectral weight along the $\overline{\Gamma}-\overline{Y}$ direction [highlighted by orange dashed circles in Fig.~\ref{fig2}(b)]. This additional intensity appears exclusively near $\overline{Y}$ and is absent around $\overline{X}$, indicating a pronounced in-plane anisotropy not observed in the bilayer analog~\cite{yang2024orbital}. Notably, a recent ARPES study of the isostructural compound La$_4$Ni$_3$O$_{10}$ recognized a small electron pocket ($\delta$ band) centered at the $\Gamma'$ point~\cite{du2025dichotomy}, while it is different from the band features observe in Pr$_4$Ni$_3$O$_{10}$.

To resolve these fine features, the sample was reoriented to align with the electron analyzer slit direction. In this experimental geometry, the enhanced momentum resolution along the vertical cut allows previously obscured band features to be clearly resolved, as marked by red arrows in Figs.~\ref{fig3}(a, b and d).

The high-resolution data reveals that the faint spectral weight is composed of multiple, distinct band crossings. Previous ARPES studies on La$_3$Ni$_2$O$_7$ have reported a small arc-like Fermi surface segment inside the $\beta$ pocket, which have been attributed to either DW band folding orders~\cite{au2025universal} or impurity doping~\cite{yang2024orbital}. Our analysis reveals that these anisotropic features in Pr$_4$Ni$_3$O$_{10}$ align remarkably well with the dispersion of the main $\beta$ band when translated in momentum space along the $\overline{\Gamma}$-$\overline{X}/\overline{Y}$ direction, as illustrated in Figs.~\ref{fig3}(c, d). This implies that these additional bands originate from band-folding driven by charge or spin density wave vectors.

To validate this scenario, we quantitatively analyzed the momentum separation between the primary $\beta$ band and these replicas [Figs.~\ref{fig3}(e-g)]. Using the feature marked by red arrow 1 as an example, we tracked the peak positions in the momentum distribution curves (MDCs) near the Fermi level [Fig.~\ref{fig3}(e)] and determined a scattering vector of $q_1 \approx$ 0.77 $q*$. This value is consistent with the reported incommensurate CDW wave vector measured by X-ray diffraction experiments, $q_{cdw} \approx$ 0.78 $q*$~\cite{samarakoon2023bootstrapped}. A consistent analysis for the features marked by red arrows 2 and 3 [Fig.~\ref{fig3}(b, d)] yields scattering vectors of $\approx$ 0.25 and 0.61 $q*$, which correspond to 1-$q_{cdw}$ and the SDW vector $q_{sdw}$ of Pr$_4$Ni$_3$O$_{10}$, respectively [Figs.~\ref{fig3}(f and g)]. More detailed analysis of those back-folded bands are presented in the Supplementary Information section II.

Taken together, this self-consistent analysis provides compelling evidence that the additional spectral features are not new intrinsic pockets (such as a $\delta$ band), but are rather replicas of the $\beta$ band resulting from scattering by the primary CDW/SDW orders. As illustrated by the composite model in Fig.~\ref{fig3}(h), the experimentally measured Fermi surface is excellently reproduced by superimposing the intrinsic bands with their replicas folded by the known SDW and CDW vectors. These results provide a comprehensive picture of the electronic structure in Pr$_4$Ni$_3$O$_{10}$, revealing that the complex Fermi surface topology is a direct consequence of coexisting spin and charge density wave orders and octahedra distortions.

\section{Density wave related gaps}

\begin{figure*}[htbp]
\centering
\includegraphics[width=14 cm]{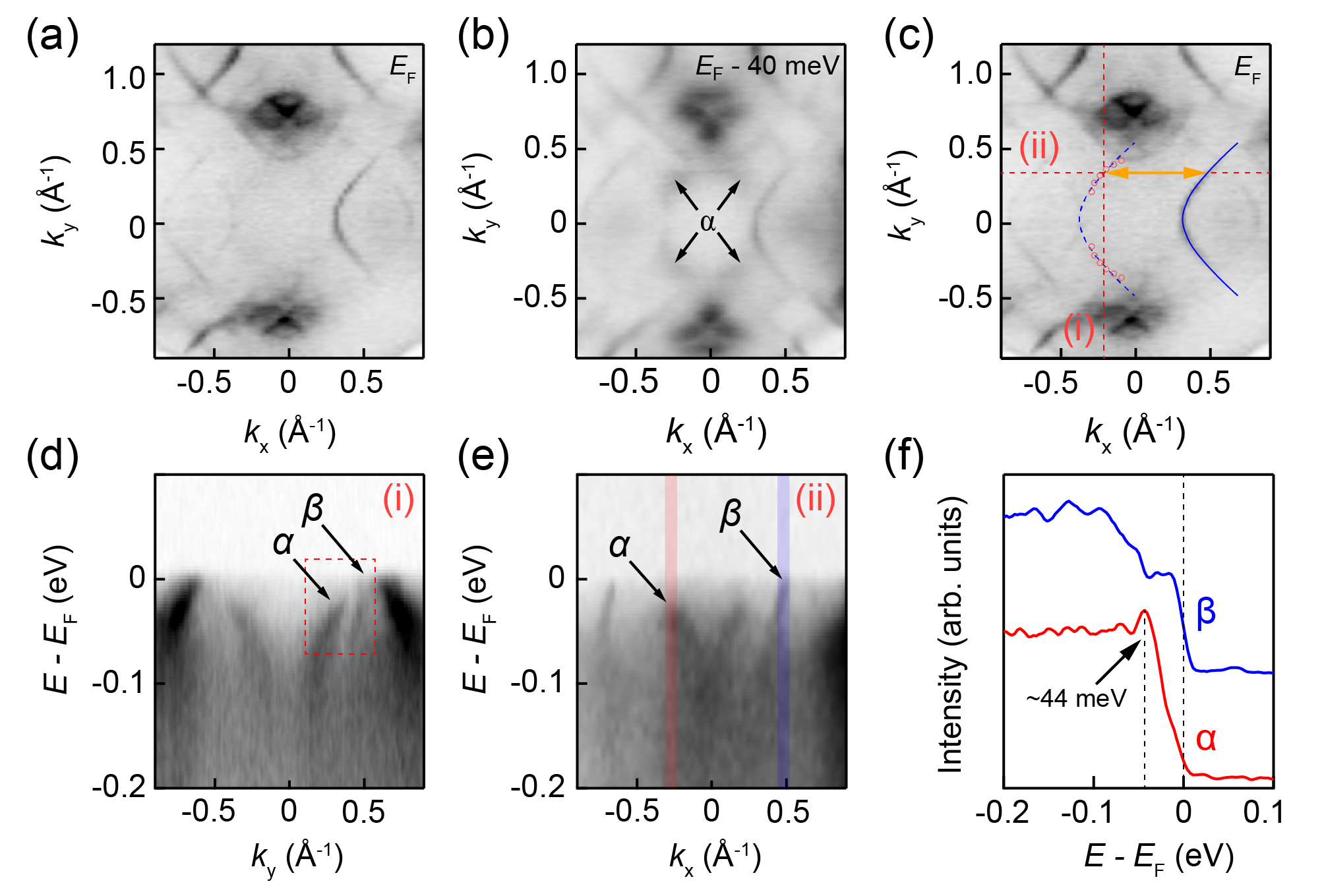}
\caption{
(a) Fermi surface intensity map at $E_F$ measured with 75 eV photons;
(b) Constant energy map at $E_F$ - 30 meV; black arrows indicate the $\alpha$ pocket.
(c) The $E_F$ - 30 meV map overlaid with the extracted $\alpha$-band counter (hollow red circles) and the $\beta$-band counter (blue lines). The dashed red lines mark the momentum cut (i) and (ii) used in (d,e); the orange arrow highlights the nesting relation between the $\alpha$ and $\beta$ pockets.
(d,e) Energy-momentum dispersions along cut (i) and (ii), respectively. 
(f) The EDCs extracted from the shaded momentum windows in (e), taken at the $\alpha$-band top (red) and at $k_F$ of the $\beta$ band (blue).
}
\label{fig4}
\end{figure*}

Prior ARPES studies on trilayer nickelates reveal orbital- and momentum-selective DW reconstruction. Spectral features observed on the hole-like $\gamma$-band~\cite{li2017fermiology} and electron-like $\alpha$-band~\cite{du2025dichotomy} suggest a sheet-dependent coupling strength. Given the multi-sheet Fermi surface topology and unidirectional ordering vectors, it is crucial to identify which specific states are coupled by the ordering vector. Our comprehensive Fermi surface mapping enables a band-resolved analysis. At low temperatures, the $\gamma$-band top lies approximately 30 meV below the Fermi level and remains stationary across $T_{DW}$ (Fig.~S6). This finding contrasts with the band shifting reported in La$_4$Ni$_3$O$_{10}$~\cite{li2017fermiology} but aligns with recent observations in La$_3$Ni$_2$O$_7$~\cite{yang2024orbital}, suggesting the $\gamma$ band is not the primary location for the DW gap.

We next turned our attention to the electron-like $\alpha$ pocket around the zone center. At 8~K, well below $T_{DW}$, the spectral weight of the $\alpha$ band is significantly suppressed at $E_F$ [Fig.~\ref{fig4}(a)]. However, in the constant energy map taken at $E_F$ - 40 meV [Fig.~\ref{fig4}(b)], the $\alpha$ band becomes clearly resolved, indicative of a sizable density-wave gaps therein. By extracting the contours of the $\alpha$ pocket (red hollow circles) from this map and superimposing them onto the Fermi surface map [Fig.~\ref{fig4}(c)], we observed that the $\alpha$ pocket coincides well with the shifted $\beta$ pocket ($\beta_{k-q_{sdw}}$). This geometric overlap provides strong evidence that Fermi surface nesting occurs between the $\alpha$ and $\beta$ bands. Note that this finding is consistent with theoretical calculations, which suggest that the inter-orbital nesting dominates over intra-orbital contributions in driving the density-wave instability in trilayer nickelates~\cite{zhang2020intertwined,jia2025lattice}. 

To quantitative analyze the density-wave gap on the $\alpha$ band, we extracted band dispersion spectra [Figs.~\ref{fig4}(d) and~\ref{fig4}(e)] along the two orthorhombic directions cutting through the $\alpha$ pocket [marked as cuts (i) and (ii) in Fig.~\ref{fig4}(c)], respectively. Both dispersions display pronounced back-bending for the $\alpha$ band, a characteristic feature of a DW-induced gap, whereas the $\beta$ band shows much weaker modification. Quantitatively, based on the corresponding EDCs [Fig.~\ref{fig4}(f)], we resolve a DW gap of approximately 44 meV in the $\alpha$ band. This magnitude is comparable to that observed by STM and optical spectroscopy in La$_4$Ni$_3$O$_{10}$~\cite{li2025direct,xu2025collapse}. Such orbital-selective gap can be well captured by the calculated band structure, as shown in SI section V. It is worth noting that the $\alpha$ band gap appears largely momentum independent, revealing a isotropic gap on $\alpha$ pocket, as shown in Fig.~S5. 

\section{Band splitting of the \texorpdfstring{$\beta$}{beta} pockets}

\begin{figure}[htbp]
\centering
\includegraphics[width=8cm]{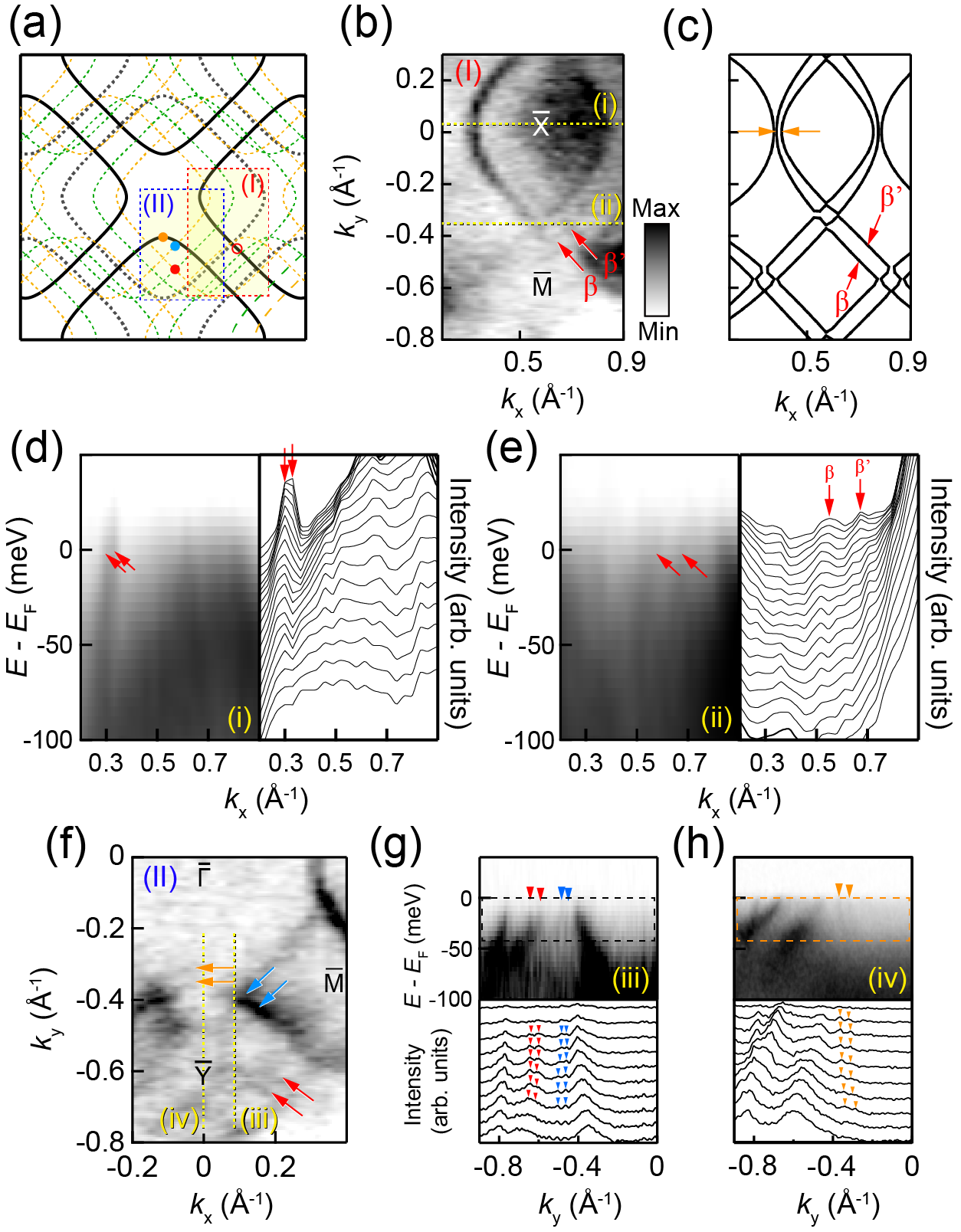}
\caption{
(a) Schematic Fermi surface including all relevant bands. The red (blue) dashed box marks region~I (II), corresponding to the zoomed Fermi-surface maps in (b) [(f)]. Colored dots indicate representative momenta $k_{\beta}$ used for the splitting analysis in (e) and (g,h).
(b) Zoomed Fermi-surface intensity map measured at 8~K with 75~eV photons.
(c) Tight-binding Fermi surface calculated with an outer--outer NiO$_2$-plane hopping $t_{\perp}^{x_{oo}}=50$~meV.
(d,e) Energy--momentum dispersions from the 75~eV dataset along the cuts $k_y=0$ and $k_y$=-0.38~\AA$^{-1}$, respectively, as indicated by the yellow dashed lines (i) and (ii) in (b); the corresponding MDC stacks are shown on the right.
(f) Zoomed Fermi-surface intensity map measured at 8~K with 49~eV photons; arrows highlight the resolved $\beta/\beta'$ splitting.
(g,h) Energy-momentum dispersions from the 49~eV dataset along the cuts $k_x=0$ and $k_x$ = 0.13~\AA$^{-1}$, respectively, as indicated by the yellow dashed lines (iii) and (iv) in (f). The accompanying MDCs (bottom panels) are integrated over the energy window from $E_F$ to $E_F-50$~meV. Inverted triangles mark the extracted band splittings.
}
\label{fig5}
\end{figure}

With the Fermi-surface complexity arising from lattice distortions and density-wave folding now fully characterized, another longstanding puzzle is why the trilayer system appears strikingly similar to its bilayer counterpart, despite the expectation that interlayer coupling within the trilayer block should induce a distinct bonding–antibonding splitting. In previous ARPES studies on La$_4$Ni$_3$O$_{10}$, this expected $\beta$-band splitting has never been resolved, and the absence was attributed to either limited experimental resolution or matrix-element suppression~\cite{li2017fermiology}.

By rigorously distinguishing intrinsic features from the band-folding artifacts discussed above, we are now able to resolve this intrinsic splitting. As illustrated in Fig.~\ref{fig5}(a), we identify two distinct regions (marked by red and blue dashed rectangular frames) where the $\beta$-band splitting is resolved using 75 eV and 49 eV photons, respectively. Fig.~\ref{fig5}(b) displays an enlarged view of the Fermi surface within Region I. There, the $\beta$ band is clearly split into two branches, denoted $\beta$ and $\beta'$, and this splitting aligns well with DFT predictions for the bonding/antibonding states [Fig.~\ref{fig1}(c)]. To better distinguish the split $\beta$ band, we extract the band dispersion spectra [Figs.~\ref{fig5}(d) and~\ref{fig5}(e) ] along the $\overline{\Gamma}$–$\overline{X}$ direction and along an off–high-symmetry direction at $k_y = -0.38~\text{\AA}^{-1}$ [cuts (i) and (ii) in Fig.~\ref{fig5}(b)], respectively. Along both directions, the $\beta$ band splits into two distinct branches, with the MDCs confirming two well-defined peaks of comparable intensity. Crucially, the location of the $\beta'$ branch does not coincide with any projection from our density-wave folding model [indicated by the red hollow circle in Fig.~\ref{fig5}(a)], ruling out a CDW/SDW origin. 

We also resolve distinct band splittings in Region II, specifically at the locations marked by red, blue and orange points in Fig.~\ref{fig5}(a). Along the off-symmetry cut (iii) illustrated in Fig.~\ref{fig5}(f), equivalent splitting features are observed in the back-folded bands, $\beta_{\mathrm{sdw}}$ (blue arrows) and $\beta_{\mathrm{cdw}}$ (red arrows). The corresponding band dispersions and MDCs across these two bands are displayed in Fig.~\ref{fig5}(g), where triangles highlight the positions of $\beta_{sdw/cdw}$ and their respective splittings.
Surprisingly, along the high-symmetry $\overline{\Gamma}$–$\overline{Y}$ direction, the $\beta$ band exhibits a sizable splitting, in clear contrast to the near degeneracy predicted by the DFT calculations shown in Fig.~\ref{fig1}(c). This pronounced splitting is consistently observed both in the energy–momentum dispersion along this direction and in the corresponding MDCs, as highlighted by the orange triangles in Fig.~\ref{fig5}(h).
From a symmetry perspective, the $\overline{\Gamma}$–$\overline{Y}$ line is a high-symmetry direction along which the $\beta$ band is composed purely of $d_{x^2-y^2}$ orbital character. Since interlayer coupling between $d_{x^2-y^2}$ orbitals is generally weak, one would expect the $\beta$ bands to remain nearly degenerate along this direction.
To identify the origin of the observed splitting, we constructed a tight-binding (TB) model (see Sec.~V of the Supplemental Information). We find that interlayer hopping between the inner and outer layers alone, characterized by $t_{\perp}^{x_{io}}$, is insufficient to reproduce the observed splitting. In contrast, introducing an additional hopping term between the $d_{x^2-y^2}$ orbitals of the top and bottom Ni atoms, denoted $t_{\perp}^{x_{oo}}$, leads to a splitting at the same momentum location as seen in the ARPES measurements. A clearly resolvable splitting emerges only when $t_{\perp}^{x_{oo}}$ reaches approximately 50~meV, as indicated by the orange arrows in Fig.~\ref{fig5}(c) and Fig.~S8.
These results point to unexpectedly complex interlayer hopping processes in trilayer nickelates, highlighting the need for further experimental and theoretical investigations.

\section{Summary}
Our comprehensive $\mu$-ARPES investigation establishes a coherent picture of how trilayer coupling, octahedral distortions, and intertwined DW orders cooperate to shape the low-energy electronic structure of Pr$_4$Ni$_3$O$_{10}$. By selecting a single structral/DW domain, we eliminate the spectral averaging that obscured prior ARPES studies of RP nickelates, allowing us to explicitly separate intrinsic dispersions from folding artifacts. This enables (i) a domain-resolved identification of SDW- and CDW-induced back-folded bands with distinct incommensurate vectors; in light of recent transport evidence suggesting a close connection between SDW order and superconductivity in the RP nickelate family~\cite{shi2025pressure,zhao2025pressure,shi2025absence}, our results highlight the importance of explicitly accounting for SDW domain selectivity in future electronic-structure studies; (ii) a momentum-resolved determination of the dominant regions for DW gap formation, which validating the existence of the gap in momentum space and identifying inter-orbital nesting as the primary driver of the SDW instability; and (iii) the direct resolution of the long-sought intrinsic trilayer $\beta$-band splitting and discovery of inter-layer hopping between the outmost Ni-O layers. Together, these findings provides a solid foundation for modeling the interplay density waves instability and electronic structures in trilayer nickelate superconductors.
\section{Acknowledgment}
This work is supported by National Natural Science Foundation of China (Grants No.12494593, No.12504079), National Key R\&D Program of China (Grants No. 2024YFA1408103, No. 2023YFA1406304), Innovation Program for Quantum Science and Technology (Grant No. 2021ZD0302803), and New Cornerstone Science Foundation. Z.C.J. acknowledges the China National Postdoctoral Program for Innovative Talents (BX20240348) and Xiaomi Young Talents Program. D.W.S. acknowledges the Anhui Provincial Natural Science Foundation (Grant No. 2408085J003). We thank the Shanghai Synchrotron Radiation Facility (SSRF) for the beamtime on beamline 03U (31124.02.SSRF.BL03U), which is supported by ME$^2$ project under contract No. 11227902 from National Natural Science Foundation of China.

\section{Author Contribution}
D.W.S. and D.L.F. proposed and designed the research. E.K.Z. and J.Z. contributed to single crystal growth and characterizations. Y.X.W and K.J. contributed to the DFT band calculations. Z.C.J., R.F.Z. and Y.H. carried out the ARPES experiments. Z.C.J., Z.T.L., J.S.L., R.F.Z., W.C.J., Y.H., Q.J., M.Y., D.W.S. contributed to the development and maintenance of ARPES system and beamline. Z.C.J., E.K.Z., Y.X.Z., Z.T.L., X.N.Z. and D.W.S. analyzed the data. Z.C.J., Z.T.L., Y.X.W., K. J. D.W.S. and D.L.F. wrote the manuscript. All authors participated in the discussion and comment on the paper.

\section{Methods}
\subsection{A. Sample growth and characterization}
Precursor powder of Pr$_{4}$Ni$_{3}$O$_{10}$ was prepared by a standard solid-state reaction. Stoichiometric cation ratios corresponding to Pr$_{4}$Ni$_{3}$O$_{10}$ were obtained by mixing Pr$_{6}$O$_{11}$ (Aladdin, 99.99 \%) and NiO (Aladdin, 99.99 \%). The powders were thoroughly ground and mixed, then calcined at 1373 K three times to ensure a complete and homogeneous reaction. An additional 0.5 \% NiO was added to compensate for possible volatilization during crystal growth.
Subsequently, the reacted powder was isostatically pressed into 12 cm-long, 6 mm-diameter rods and sintered in air at 1673 K for 12 h. Single crystals were grown in a vertical optical floating-zone furnace (Model HKZ, SciDre) via a two-step procedure: first, the rod was rapidly scanned (20 mm h$^{-1}$, 10 bar O$_{2}$) to enhance density; then growth was carried out at 2 mm h$^{-1}$ under 130–150 bar O$_{2}$.

Transport measurements were performed in a Quantum Design PPMS (1.8–300 K) using a standard four-probe method. Gold pads were deposited on the sample first to ensure stable connections during measurement.

\subsection{B. ARPES Measurements}
Synchrotron-based ARPES measurements were carried out at the BL03U endstation of the Shanghai Synchrotron Radiation Facility (SSRF), using a Scienta DA30 electron analyzer with vertical slit of 300 $\mu$m (perpendicular to the laboratory floor). The base pressure during measurements was lower than 8 $\times$ 10$^{-11}$ Torr. All Samples were cleaved $\textit{in suit}$ below 10~K at in the measurement chamber. The measurement temperature was maintained at 8~K unless otherwise specified in the main text. A focused beam spot smaller than 15 $\times$ 15 $\mu$m$^2$ was adopted during the measurements. The measurements were mainly performed with photon energies of 49 and 75 eV with linear horizontal (LH) light (parallel to the laboratory floor). The overall energy resolution was 8$\sim$10 meV and angular resolution was $\sim 0.3^{\circ}$.

\subsection{C. Calculations}
Our DFT calculations employ the Vienna ab-initio simulation package (VASP) code \cite{kresse1996efficient} with the projector augmented wave (PAW) method \cite{kresse1999ultrasoft}. The exchange-correlation functional employs the SCAN functional \cite{PhysRevLett.115.036402}, a type of meta-GGA approach that offers higher accuracy than traditional GGA functionals. The cutoff energy for expanding the wave functions into a plane-wave basis is set to be 500 eV. The energy convergence criterion is 10$^{-8}$
eV. All calculations are conducted using the primitive cell. The $\Gamma$-centered 9$\times$9$\times$4 k-meshes are used. When calculating the Fermi surface, first apply the Wannier90 software \cite{Pizzi_2020} to fit the DFT-calculated bands, then use the generated hr file to compute the Fermi surface. Directly modifying the hopping parameters in the hr file allows for obtaining Fermi surfaces under different parameters.
\

\bibliographystyle{naturemag}
\bibliography{PNO}
\end{document}